\documentclass[prd,showpacs,floatfix,preprint]{revtex4}
\usepackage{amsmath,amsfonts,amssymb}
\usepackage[dvips]{graphicx}

\usepackage{amsmath}

\begin{document}

\title{Effects of the quantization ambiguities on the Big Bounce dynamics}

\author{Orest Hrycyna} 
\email{hrycyna@kul.lublin.pl}
\affiliation{Department of Theoretical Physics, Faculty of Philosophy, 
The John Paul II Catholic University of Lublin, Al. Rac{\l}awickie 14, 20-950
Lublin, Poland}

\author{Jakub Mielczarek}
\email{jakubm@poczta.onet.pl}
\affiliation{Astronomical Observatory, Jagiellonian University, 30-244
Krak\'ow, Orla 171, Poland}
\affiliation{Institute of Physics, Jagiellonian University, Reymonta 4, 30-059 
Krak\'ow, Poland}

\author{Marek Szyd{\l}owski}
\email{uoszydlo@cyf-kr.edu.pl}
\affiliation{Astronomical Observatory, Jagiellonian University, 30-244
Krak\'ow, Orla 171, Poland}
\affiliation{Marc Kac Complex Systems Research Centre, Jagiellonian University,
Reymonta 4, 30-059 Krak{\'o}w, Poland}

\date{\today}

\begin{abstract}
In this paper we investigate dynamics of the modified loop quantum cosmology 
models using dynamical systems methods. Modifications considered come from the 
choice of the different field strength operator $\hat{F}$ and result in 
different forms of the effective Hamiltonian. Such an ambiguity of the choice 
of this expression from some class of functions is allowed in the framework of 
loop quantization. Our main goal is to show how such modifications can 
influence the bouncing universe scenario in the loop quantum cosmology. In 
effective models considered we classify all evolutional paths for all 
admissible initial conditions. The dynamics is reduced to the form of a 
dynamical system of the Newtonian type on a 2-dimensional phase plane. These 
models are equivalent dynamically to the FRW models with the decaying effective 
cosmological term parameterized by the canonical variable $p$ (or by the scale 
factor $a$). We demonstrate that the evolutional scenario depends on the 
geometrical constant parameter $\Lambda$ as well as the model parameter $n$. 
We find that for the positive cosmological constant there is a class of 
oscillating models without the initial and final singularities. The new 
phenomenon is the appearance of curvature singularities for the finite values 
of the scale factor, but we find that for the positive cosmological constant
these singularities can be avoided. The values of the parameter $n$ and the 
cosmological constant differentiate asymptotic states of the evolution. For the 
positive cosmological constant the evolution begins at the asymptotic state in 
the past represented by the de Sitter contracting (deS$_{-}$) spacetime or 
the static Einstein universe $H=0$ or $H=-\infty$ state and reaches the
de Sitter expanding state (deS$_{+}$), the state $H=0$ or $H=+\infty$ state. In
the case of the negative cosmological constant we obtain the past and future
asymptotic states as the Einstein static universes.
\end{abstract}

\maketitle

\section{Introduction} 
\label{Intro}

According to the General Relativity the initial state of the universe was 
a causal singularity. In this point all time-like geodesics were cut off and 
has no extension into the past. This peculiar behaviour corresponds to the 
blow up of the curvature $\mathcal{R} \rightarrow \infty$ and energy density 
$\rho \rightarrow \infty$  in the standard Big-Bang scenario. It is however 
believed that singularities of this kind are not real states but rather 
artifacts resulting from the fact that classical theory broke down. In the 
vicinity of the initial singularity, when the universe reaches the Planck scale 
energy, the quantum gravity effects were proposed as a mechanism to remove 
the singular behaviour. Such a possibility was studied in the context of the 
String Theory leading to the so called pre-Big Bang cosmological scenario 
\cite{Gasperini:2007vw}. Presently this issue is also studied extensively 
in the Loop Quantum Cosmology (LQC)\cite{Bojowald:2006da} where the avoidance 
of the initial singularity appeared in a natural way \cite{Bojowald:2007ky}. 
This approach possesses other advantages like the adjustment of the proper 
initial conditions for the inflationary phase. The avoidance of the initial 
singularity appeared here in the form of the non-singular bouncing universe 
\cite{Ashtekar:2006rx,Bojowald:2006gr}. Namely, the universe is initially in 
the contracting phase, then reaches a minimal nonzero volume and due to
quantum repulsion evolves toward an expanding phase. It should be stressed that
this result has been strictly demonstrated so far only for 
isotropic models with a free, massless scalar field \cite{Ashtekar:2006wn}. 
The phenomenological dynamics of this 
scenario has been extensively studied \cite{Singh:2006im, Mielczarek:2008zv}. 
However ambiguities in the quantization of the classical formula lead to some 
deviations. In the frame of the standard quantization scheme these ambiguities 
lead generally to the models with qualitatively similar bouncing behaviour. 
However like it was shown in \cite{Mielczarek:2008zv}, for some special setup 
in the effective flat FRW model the bounce is replaced by non-singular 
oscillations. More drastic consequences appear when we choose the different 
quantization framework. This issue was studied in \cite{Mielczarek:2008mf} 
and it was found the existence of the finite scale factor curvature 
singularities in the co-existence with a quantum bounce. In this paper we 
continue these studies in the nonstandard quantization frameworks. We 
investigate the effects of resulting ambiguities for the dynamics of the 
effective FRW models with the cosmological constant. We present simple methods 
of reducing dynamical equation governing evolution of the models with quantum 
corrections to the form of a 2-dimensional autonomous dynamical system of the 
Newtonian type. We explore a mechanical analogy in investigation of different 
evolutionary scenarios.  

The organization of the text is the following. In section \ref{sec:2}
we investigate possible kinds of ambiguities in the loop quantization,
concentrating on the expression for the field strength operator $\hat{F}$.
Next, in section \ref{sec:3} we investigate resulting cosmological models with
use of the qualitative methods analysis of dynamics.
In section \ref{sec:summary} we summarize the results.

\section{Field strength operator $\hat{F}$ and quantization ambiguities}
\label{sec:2}

This section aims to summarize some key features of loop quantization of 
cosmological models and discusses arising ambiguities. In particular we extend 
study of the choice of the field strength operator $\hat{F}$ began in 
\cite{Mielczarek:2008mf}. We restrict our consideration to the flat FRW 
cosmological models which are given by the metric
\begin{equation}
ds^2=-N^2(x) dt^2 + q_{ab}dx^adx^b
\end{equation}
where $N(x)$ is the lapse function and the spatial part of the metric is expressed as 
\begin{equation}
q_{ab}= \delta_{ij} {\omega^i_a} {\omega^j_b}= a^2(t) {^oq}_{ab} = a^2(t)  \delta_{ij}  {^o\omega^i_a}{^o\omega^j_b}.
\end{equation}
In this expression ${^oq}_{ab}$ is the fiducial metric and ${^o\omega^i_a}$ are the co-triads dual to the triads 
${^oe^a_i}$,  ${^o\omega^i}({^oe_j})=\delta^i_j$ where $^o\omega^i={^o\omega^i_a}dx^a$ and $^oe_i={^oe_i^a}\partial_a$. 
Based on the triads  we can construct the Ashtekar variables 
\begin{eqnarray}
A^i_a &\equiv& \Gamma^i_a+\gamma K_a^i = c V_0^{-1/3} \ {^o\omega^i_a}  , \label{A} \\
E^a_i &\equiv& \sqrt{|\det q|} e^{a}_i = p V_0^{-2/3} \sqrt{^oq} \ {^oe^a_i} \label{E}
\end{eqnarray}  
where 
\begin{eqnarray}
|p| &=& a^2 V_0^{2/3}, \\
 c &=& \gamma \dot{a} V_0^{1/3}.
\end{eqnarray}
The volume $V_0$ is the so called fiducial cell and is introduced to perform 
the integration on the non-compact spaces. We expect that the physical results
should not depend on its value. However, when we introduce quantum
modifications in general they can depend on the choice of $V_{0}$.

In loop quantum cosmology we consider effects of the inverse volume corrections
and holonomies. In the phenomenological considerations of the quantized FRW $k=0$ model it is 
appropriate to take into account quantum modifications arising only from the 
field strength expressed in terms of holonomies. In some special case ($n=-1/2$) this kind
of modification does not depend on the choice of the fiducial cell $V_{0}$. This
simplifies interpretation of results and make the correspondence with 
classical physics easier. This issue
is related to the question of the loop quantization uniqueness of
 the cosmological models \cite{Corichi:2008zb}.
Because there is no defined
length scale in the flat FRW models the $V_{0}$ dependence seems to be
unphysical. This problem is not present in the FRW 
$k=\pm 1$ models, where the length scale is tied with the intrinsic curvature
\cite{Ashtekar:2006es}. 

The loop quantization requires to replace the classical variables by the 
quantum operators expressed in terms of the holonomies and fluxes. In the FRW 
models holonomies reduce to almost periodic functions 
$\sum_{j} \xi_j e^{i \mu_j c/2 }$ and fluxes reduce to variable $p$ multiplied 
by the kinematic factor \cite{Ashtekar:2003hd}. Then algebra of variables $p$ 
and $e^{i \mu c/2 }$ is quantized. The quantum Poisson bracket of elementary 
operators is 
\begin{equation}
\left[\hat{p},\widehat{e^{\frac{i\mu c}{2}}} \right]  =\mu \frac{8\pi l_{\text{Pl}}^2 \gamma} {6} \widehat{e^{\frac{i\mu c}{2}}}
\end{equation}
and these operators act in the Hilbert space 
$L^2(\mathbb{R}_{\text{Bohr}}, d\mu_{\text{Bohr}} )$. The basis of this space 
can be chosen as a set of eigenstates of the operator $\hat{p}$ 
\begin{equation}
\hat{p}| \mu \rangle  =\mu \frac{8\pi \gamma l^2_{\text{Pl}}}{6}  | \mu \rangle
\end{equation}
which fulfils the normalization condition $\langle \mu | \mu' \rangle = \delta_{\mu,\mu'} $.
The action of the operator $\hat{N} = \widehat{e^{\frac{i\mu c}{2}}} $ is following 
\begin{equation}
\widehat{e^{\frac{i\mu c}{2}}} | \mu' \rangle =|\mu'+\mu \rangle.
\end{equation}
 
We have defined a quantum kinematic sector of the loop quantum cosmology. Now with 
use of these rules we aim to quantize the scalar constraint 
\begin{eqnarray}
\mathcal{H}_{\rm G} = \frac{1}{16 \pi G} \int_{\Sigma} d^3 x N(x) \frac{E^a_i
E^b_j}{\sqrt{|\mathrm{det} E|}}  \left[  {\varepsilon^{ij}}_k F_{ab}^k-2(1+\gamma^2)  K^i_{[a}  K^j_{b]} \right]
\label{scalar}
\end{eqnarray}
where the extrinsic curvature is defined as 
\begin{equation}
K_{ab}=\frac{1}{2N}\left[ \dot{q}_{ab} -2 D_{(a}N_{b)} \right]
\end{equation}
which corresponds to $K^i_a := K_{ab} e^b_i$. The scalar constraint quantum operator 
should be expressed in terms of the elementary operators $\hat{p}$ and $\hat{N} = \widehat{e^{\frac{i\mu c}{2}}} $.
In our considerations we restrict ourselves to
quantization of the field strength $F_{ab}^k$. The rest of the terms lead to 
quantum operators expressed in term of the inverse volume operator. 
Detailed analysis of the scalar constraint quantization can be found in \cite{Vandersloot:2005kh}. 

As we have said our main interest is to find a formula for the field strength 
operator $\hat{F}$. The standard approach to introduce this operator is to 
consider a closed path $\alpha = \Box_{ij} \in \Sigma$ and holonomy along this 
loop. Such a holonomy can be written as 
\begin{equation}
h_{\Box_{ij}}^{(\mu)} = h_{i}^{(\mu)} h_{j}^{(\mu)} h_{i}^{(\mu)-1} h_{j}^{(\mu)-1} 
\label{holo1}
\end{equation}
where particular holonomies are 
\begin{equation}
h_{i}^{(\mu)} = e^{\tau_i \mu c } 
              = \mathbb{I}\cos \left( \frac{\mu c}{2}\right)+2\tau_i\sin \left( \frac{\mu c}{2}\right).
\label{hol2}
\end{equation}
With use of the above expressions one can derive formula
\begin{equation}
F^k_{ab} =  - 2 
 \frac{\text{tr}\left[\tau_k \left( h^{(\mu)}_{\Box_{ij}}-\mathbb{I} \right)  \right]}{\mu^2 V_0^{2/3} } 
{^o\omega^i_a}{^o\omega^j_b}  +\frac{\mathcal{O}(\mu^4)}{\mu^2}.
\label{classF}
\end{equation}

Based on this expression we can define the quantum field strength operator. 
However, we should discuss some related issues. Firstly, quantization of the 
area operator in the Loop Quantum Gravity leads to the existence of the area 
gap $\text{Ar}_{\text{min}} = \Delta$ \cite{Ashtekar:1996eg}. This means that 
the limit $\text{Ar}_{\Box_{ij}} \rightarrow 0$ of the area enclosed by the 
loop does not exist. We should stop shrinking the loop at a minimal area which 
is related to limiting $\mu \rightarrow \bar{\mu}$ value. This process is 
however ambiguous and generally limiting $\mu$ can have a form 
\begin{equation} 
\bar{\mu} = \xi |p|^n
\end{equation}
where $n \in [-1/2 ,0)$  \cite{Bojowald:2008pu} and $\xi$ is
dimension-full and positive real parameter. 
In particular, from this expression we can recover known $\bar{\mu}$-scheme 
for which 
\begin{equation} 
\bar{\mu} = \sqrt{\frac{\Delta}{|p|}}.
\end{equation}
In this case the area gap $\Delta$ corresponds to the physical area of the 
loop $|p| \bar{\mu}^2$. The range of the $n$ parameter written above can be 
obtained from the considerations of the lattice states \cite{Bojowald:2006qu}.
Further restrictions for value of this parameter come from considerations of 
the anomaly cancellation in the inhomogeneous cosmological models and 
positivity of the graviton effective mass, giving $-0.1319 > n \geq -5/2$ 
\cite{Bojowald:2007cd}.  

The properties considered should be now taken into account during the construction 
of the operator $\hat{F}$. The standard procedure is to cut-off the terms 
$\mathcal{O}(\mu^4)$ and perform the limit $\text{Ar} \rightarrow \Delta$ 
in expression (\ref{classF}), what gives \cite{Ashtekar:2006uz}
\begin{equation}
\hat{F}^k_{ab} = - 2 
\lim_{\text{Ar} \rightarrow \Delta}
\frac{\text{tr}\left[\tau_k \left( h^{(\mu)}_{\Box_{ij}}-\mathbb{I} \right)  \right]}{\mu^2 V_0^{2/3} } 
{^o\omega^i_a}{^o\omega^j_b} =
\frac{  \sin^2\left(\bar{\mu} c \right) }{ \bar{\mu}^2 V_0^{2/3} } \epsilon_{kij}  {^o\omega^i_a}{^o\omega^j_b}.
\label{op1}
\end{equation}
This is the well defined operator on the Hilbert space 
$L^2(\mathbb{R}_{\text{Bohr}} ,d\mu_{\text{Bohr}} )$. 
Namely, due to relation $\sin\left(\bar{\mu} c \right) =\frac{1}{2i} \left( e^{i\bar{\mu} c}- e^{-i\bar{\mu} c}\right)$
it can be expressed in terms of the operator $\hat{N}$ and due to the power-low 
$\bar{\mu}(p)$ dependence this part can be expressed in terms of the operator 
$\hat{p}$. The proper classical behaviour is also obtained, it means that no 
information about classical expression is contained in the terms 
$\mathcal{O}(\mu^4)$. These terms simply vanish in the classical limit 
$\mu \rightarrow 0$, $\mathcal{O}(\mu^4)\rightarrow 0$.
Another motivation to consider (\ref{op1}) as a proper quantum operator is the 
resulting dynamics. Namely, this quantum operator leads to the avoidance of the 
initial singularity.

However we can ask what happens when a part of the $\mathcal{O}(\mu^4)$ contribute
to the definition of the operator $\hat{F}$? 
The resulting operator has proper classical behaviour, because the contribution 
from $\mathcal{O}(\mu^4)$ vanishes in the classical limit. However on the 
quantum level the operator $\hat{F}$ contain some new terms which influence 
quantum dynamics. It is worth to stress this property. This additional 
modification term, which influence dynamics in the quantum regime, does not 
correspond to any contribution on the classical level. The possibility of 
existence of such terms lead to the ambiguity in the definition of the operator 
$\hat{F}$. However not all modifications which fulfil well the classical limit 
are allowed. Another property which must hold is that the resulting operator 
must be defined on the Hilbert space $L^2(\mathbb{R}_{\text{Bohr}}, d\mu_{\text{Bohr}} )$.

In the paper \cite{Mielczarek:2008mf} the expression for the $\hat{F}$ operator
including the term $\mathcal{O}(\mu^4)$ was calculated 
\begin{equation}
\hat{F}_{ab}^k = \epsilon^k_{\ ij}\frac{1-\sqrt{1-\frac{4}{3}\sin^2\left(\bar{\mu} c \right) }}
{\frac{2}{3}\bar{\mu}^2 V_0^{2/3}} 
{^o\omega^i_a}{^o\omega^j_b}.
\label{EffFS}
\end{equation}
Expanding the square we obtain 
\begin{equation}
\frac{1-\sqrt{1-\frac{4}{3}\sin^2\left(\bar{\mu} c \right) }}{\frac{2}{3}\bar{\mu}^2}
= \left[ \frac{\sin (\bar{\mu} c)}{\bar{\mu}}  \right]^2 +\frac{1}{3}\frac{\sin^4 (\bar{\mu} c)}{\bar{\mu}^2} +\dots. 
\label{expansion}
\end{equation}
We see that the first term in this series corresponds to the standard field 
strength operator. Other parts vanish in the limit $\bar{\mu}\rightarrow 0$ and 
have no classical meaning. With use of the formula 
$\sin\left(\bar{\mu} c \right) =\frac{1}{2i} \left( e^{i\bar{\mu} c}- e^{-i\bar{\mu} c}\right)$
we can express the series in terms of the operator $\hat{N}$.

In the next section we apply the modified expression for the field strength 
operators investigated above in the cosmological models. Our goal is to find 
whether such modifications lead to singular behaviour and how the bouncing 
universe scenario is modified. We will restrict to the flat FRW models with the 
free scalar field and the cosmological constant content. Our analysis 
concentrates on the investigation of the effective equations rather than strict 
quantum dynamics. So we obtain information only about evolutions of the mean 
values of the canonical parameters. The information about dispersions is 
unavailable in such an approach. We will consider both full modifications from 
the term $\mathcal{O}(\mu^4)$ as well quadratic modifications from the 
expression (\ref{expansion}).

\section{Qualitative analysis of dynamics}
\label{sec:3}

The dynamical equations resulting from loop quantum cosmology corrections cannot 
be integrated in the exact form and that is why the qualitative methods of 
differential equations can provide us with a powerful tool for investigations 
of evolutional paths of the models under considerations.

The main advantage of using these methods lies in possibility of examine of all
solutions for all admissible initial conditions in the geometric language of a
phase space (space of all states at any time). As a result we obtain description
of global dynamics of these models, i.e., the phase portrait from which we obtain
their clear stability analysis. In the dynamical systems framework the crucial
role plays the critical points (for which right-hand sides of dynamical system
vanish) and the trajectories joining them. From the physical point of view the
critical points represent stationary states of the system or asymptotic states. 
On the other hand trajectories represent evolution of the system. The famous 
Hartman-Grobman theorem \cite{Perko:1991} tell us that linearization of the 
system at this critical points (non-degenerate) contain all necessary
information about its behaviour in neighbourhood of this critical point. From
the linearization matrix we obtain all necessary information concerning the 
type of the critical point as well as its stability.

The dynamics of cosmological models under considerations can be reduced to the 
2-dimensional dynamical systems of the Newtonian type with the total energy $E$ 
and the potential $V$. As it is well known \cite{Perko:1991} for such systems 
only two types of critical points are admissible, namely a saddle (if 
eigenvalues of the linearization matrix are real of opposite signs) or a centre 
(if eigenvalues of the linearization matrix are purely imaginary and conjugate). 
While the first type of critical points is structurally stable, the second one 
is structurally unstable, i.e., small changes of its right-hand sides disturb 
the global phase portrait which is defined modulo a homeomorphism which
establishes the equivalence of the global dynamics. Of course the motion of the 
system in the configuration space takes place in the domain admissible for motion
determined by condition of non-negativity kinetic energy form, $E-V$ is 
non-negative. The critical points of the system are always located at the extrema 
of a diagram of the potential function (if we have an inflection point on the 
diagram then the corresponding point is degenerated). If in the diagram of the 
potential function we have local maxima then they correspond to saddles on the 
phase portrait. In turn if we have local minima on the diagram then in the phase 
space the centre type of the critical point is present. Note that for systems 
under consideration all critical points in the phase space must lie on the $p$
axis and they represent static Einstein universes. Apart from the critical
points at 
the finite domain there are present physically interesting critical points at 
infinity. Therefore for the full study of dynamics of the system the analysis 
of behaviour of trajectories at infinity is necessary. For this aims we complete 
the plane by adding to it a circle at infinity via the Poincar{\'e} sphere 
construction or equivalently by introducing the projective coordinates 
(the construction of the projective plane) \cite{Perko:1991}. In our case on
the circle at infinity critical points are located which represents the 
de Sitter universes if $n=-1/2$.

Due to the particle-like description of dynamics in the notion of classical
mechanics we obtain a simple description of evolution in terms of particle moving
in the potential well. We can classify all evolutionary pats in both
configuration and phase spaces. It is important that the dynamics is fully
characterized by a single potential function $V(p)$. If we have the global 
phase portraits of the cosmological models then one can discuss how different 
models with a desired property, say a singularity, bounce, acceleration are 
distributed in the ensemble. Are they typical or generic? The answer for these 
questions are interesting in cosmological context when problem of initial 
conditions is crucial.

\subsection{$\mathcal{O}(\mu^{4})$ modifications}

In the case of $\mathcal{O}(\mu^{4})$ modifications the phenomenological
Hamiltonian takes a form 
\begin{equation}
\mathcal{H}_{\text{phen}} = -\frac{9}{2\kappa\gamma^{2}\bar{\mu}^{2}} \Bigg( 1 -
\sqrt{1-\frac{4}{3}\sin^{2}{(\bar{\mu}c)}}\Bigg) \sqrt{|p|} + |p|^{3/2}\rho \label{HamPhen1}
\end{equation}
where matter part contains the free scalar field and the cosmological constant
\begin{equation}
\rho = \frac{p_{\phi}^{2}}{2|p|^{3}} +\frac{\Lambda}{\kappa}.
\label{rho}
\end{equation}
The Hamiltonian \ref{HamPhen1} traces the effects of the $\mathcal{O}(\mu^{4})$ modifications.
However it must be emphasized that it is not necessarily the true effective Hamiltonian.
It mean that solutions of the Hamilton equation  $\dot{f} = \left\{f,\mathcal{H}_{\text{phen}}  \right\} $
not necessarily traces the mean values of the quantum operator $\langle \hat{f} \rangle$. 
This issue must be verified by the full quantum treatment. 

From the Hamiltonian equation we obtain evolution of the canonical variable $p$ 
\begin{equation}
\dot{p} = \left\{p,\mathcal{H}_{\text{phen}}  \right\}   = \frac{2\sqrt{|p|}}{\gamma\bar{\mu}}
\frac{\sin{(\bar{\mu}c)}\cos{(\bar{\mu}c)}}
{\sqrt{1-\frac{4}{3}\sin^{2}{(\bar{\mu}c)}}}.
\end{equation}
Based on this result and the Hamiltonian constraint $\mathcal{H}_{\text{phen}}=0$ 
one derives the modified Friedmann equation in the form
\begin{equation}
H^2 \equiv \frac{1}{4}\frac{\dot{p}^{2}}{p^{2}} = \frac{\kappa}{3}\rho \left(1-\frac{\rho}{3\rho_{\text{c}}}  \right)
\left[\frac{3}{4}+\frac{1}{4}\frac{1}{\left(1-\frac{2}{3}\frac{\rho}{\rho_{\text{c}}}  \right)^2 }  \right]
\label{Fried1}
\end{equation}
where 
\begin{equation}
\rho_{\text{c}} = \frac{3}{\kappa \gamma^2 \bar{\mu}^2 |p|}.
\end{equation}
Now we make the following change of the variable $x=p^{1-n}$ and rewrite 
equation (\ref{Fried1}) to the form 
\begin{equation}
\frac{\dot{x}^2}{2} \frac{1}{(1-n)^2}\frac{3}{\kappa \rho}
 \frac{\left(1-\frac{2}{3}\frac{\rho}{\rho_{\text{c}}}  \right)^2 }
{1+3\left(1-\frac{2}{3}\frac{\rho}{\rho_{\text{c}}}  \right)^2} 
-\frac{1}{2}x^{2}(1-\frac{1}{3}\frac{\Lambda}{\Lambda_0} x^{\frac{1+2n}{1-n}})
 =-\frac{1}{2}\frac{\alpha}{\Lambda_0}                 
\end{equation} 
where $\alpha=\frac{\kappa}{6}p_{\phi}^{2}$ and $\Lambda_0 = \frac{3}{\gamma^2 \xi^2}$.
Due to the symmetry of the change $p \rightarrow -p $ we can consider
only positive branch  of $p$, as it was done above. 
In such a approach negative solutions can by simply obtained multiplying positive solutions
by minus one. 
Performing the time transformation 
\begin{equation}
\frac{d}{d\tau} =  \sqrt{\frac{1}{(1-n)^2}\frac{3}{\kappa \rho}
 \frac{\left(1-\frac{2}{3}\frac{\rho}{\rho_{\text{c}}}  \right)^2 }
{1+3\left(1-\frac{2}{3}\frac{\rho}{\rho_{\text{c}}}  \right)^2}}
\frac{d}{dt}
\label{timea}
\end{equation}
and introducing the potential function is in the form
\begin{equation}
V(x)=-\frac{1}{2}x^{2}(1-\frac{1}{3}\frac{\Lambda}{\Lambda_0} x^{\frac{1+2n}{1-n}})
\label{pota}
\end{equation}
and define 
\begin{equation}
E = -\frac{1}{2}\frac{\alpha}{\Lambda_0}
\end{equation}
we rewrite the modified Friedmann equation in the point particle form 
\begin{equation}
\frac{1}{2}x'^{2} + V(x) = E.
\label{newta}
\end{equation}

\begin{figure}
\includegraphics[scale=1]{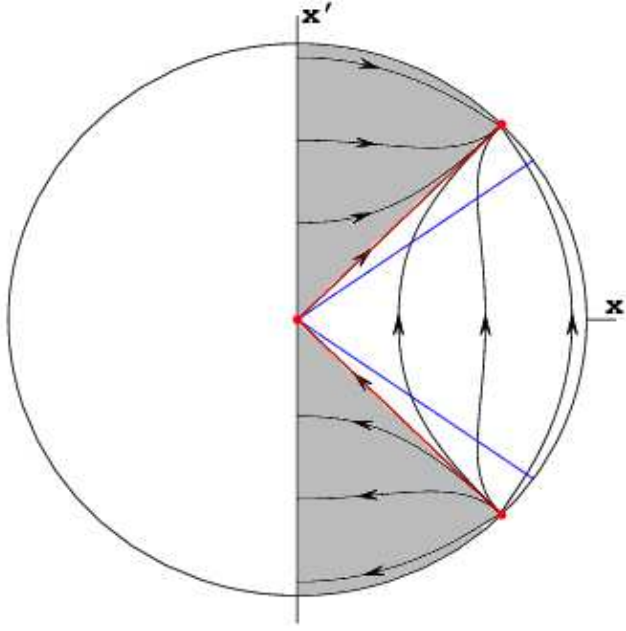}
\caption{The phase portrait for $\Lambda=0$ and arbitrary $n$ in the case A type
modifications. The shaded region
is physically forbidden. The dark gray line (blue el.v.) denotes phase space 
location of $H^{2}=\infty$. The light gray line (red el.v.) which is a border 
of the physically allowed region, denotes phase space location of $H^{2}=0$. 
The zero energy level trajectories joining critical point at the finite domain 
with critical points at infinity divides the phase space in to two region 
occupied by the scalar field ($\alpha>0$) and the phantom scalar field
($\alpha<0$) trajectories.
Only trajectories with $\alpha>0$ are physical and for all admissible initial
conditions they tend to the final state of $H^{2}=0$ for $x=\infty$ ($p=\infty$
or $a=\infty$).}
\label{fig:1}
\end{figure}

\begin{figure}
a)\includegraphics[scale=1]{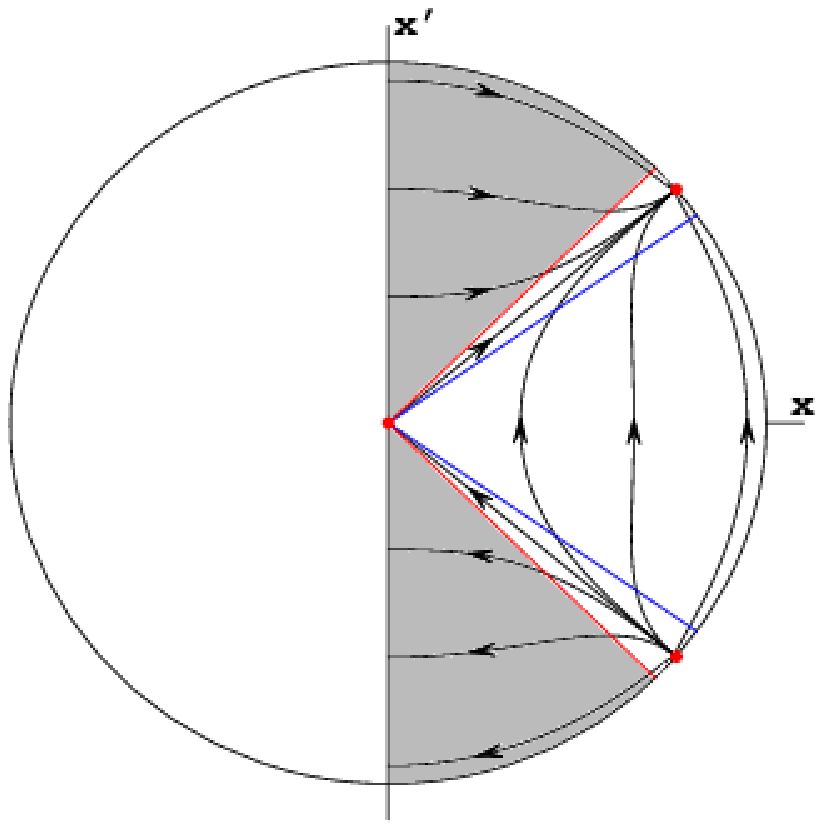}
b)\includegraphics[scale=1]{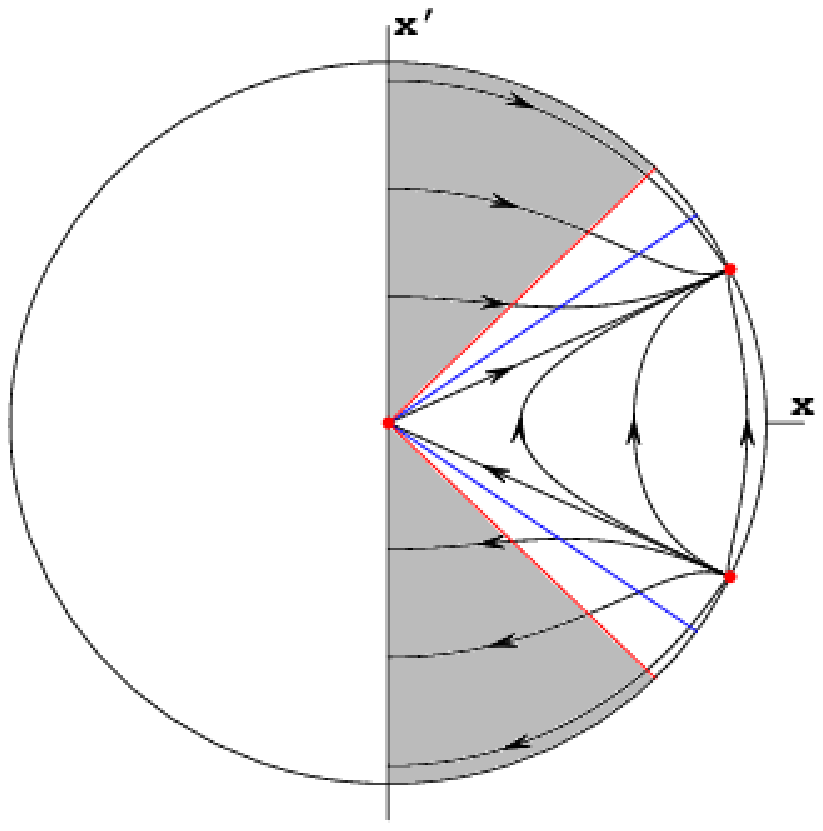}
\caption{The phase portraits for $n=-1/2$ in the case A type modifications. 
a) For $0<\frac{\Lambda}{\Lambda_0}<\frac{3}{2}$ the scalar field
($\alpha>0$) and the phantom scalar field ($\alpha<0$) trajectories are
physically allowed,
but their evolution is different; the $\alpha>0$ trajectories start from anti
de Sitter state and go through  hyper-deflation state then bounce
appears and next we have hyper-inflation state, the final state is de Sitter;
the phantom trajectories for $x'>0$ start from $H^{2}=0$ and tend to
de Sitter state;
b) For $\frac{3}{2}< \frac{\Lambda}{\Lambda_0} < 3$ the situation is quite similar
but the state of hyper-inflation (or hyper-deflation) appears only for
phantom trajectories.}
\label{fig:2}
\end{figure}

\begin{figure}
a)\includegraphics[scale=1]{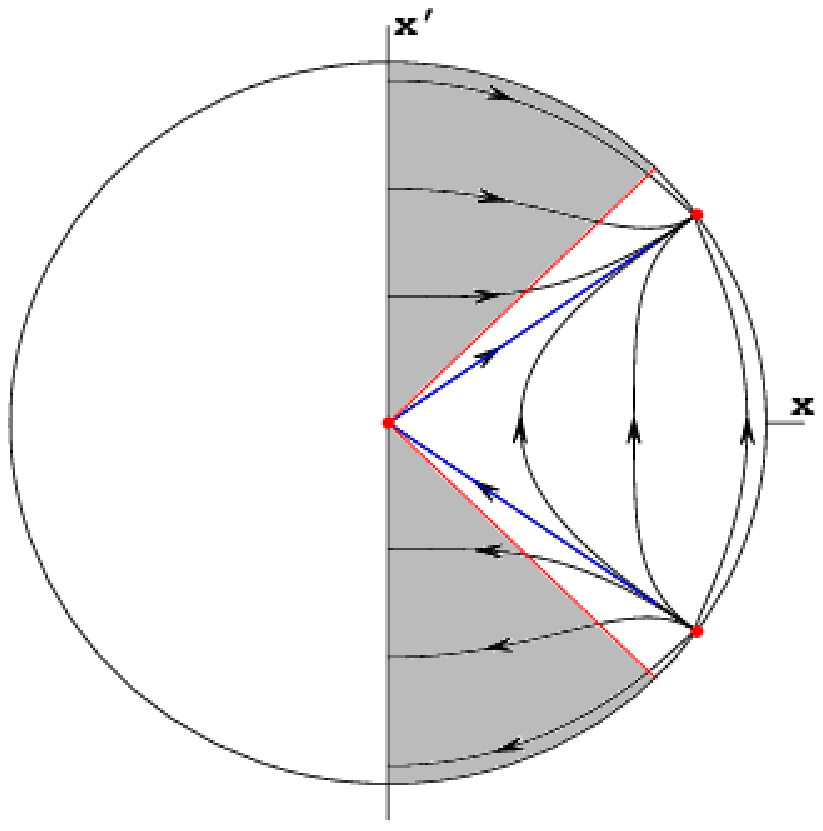}
b)\includegraphics[scale=1]{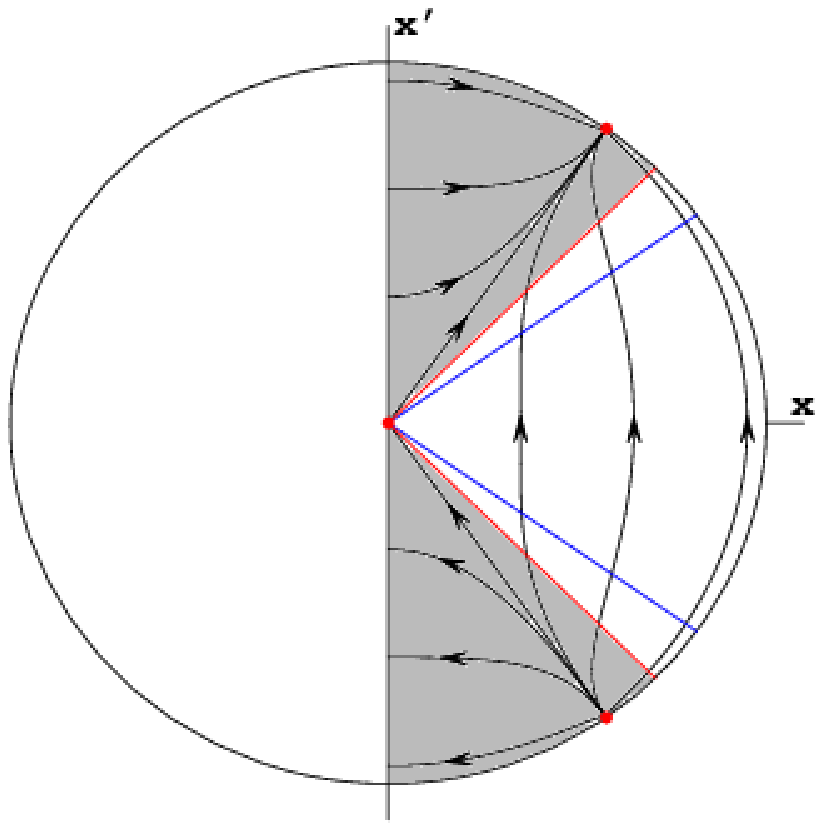}
\caption{The phase portraits for $n=-1/2$ in the case A type modifications:
a) For $\frac{\Lambda}{\Lambda_0}=\frac{3}{2}$ the trajectories with $\alpha>0$
after bounce tend to the critical state of $H^{2}=\infty$ for $x=\infty$, the
phantom ($\alpha<0$) evolution for $x'>0$ starts at $H^{2}=0$ state and
tends to the same final state of $H^{2}=\infty$; b) For $\Lambda<0$ only the
scalar field trajectories are physically allowed ($\alpha>0$), their evolution
starts at $H^{2}=0$ and after hyper-deflation, bounce and hyper-inflation
state ends at $H^{2}=0$.}
\label{fig:3}
\end{figure}

\begin{figure}
a)\includegraphics[scale=0.9]{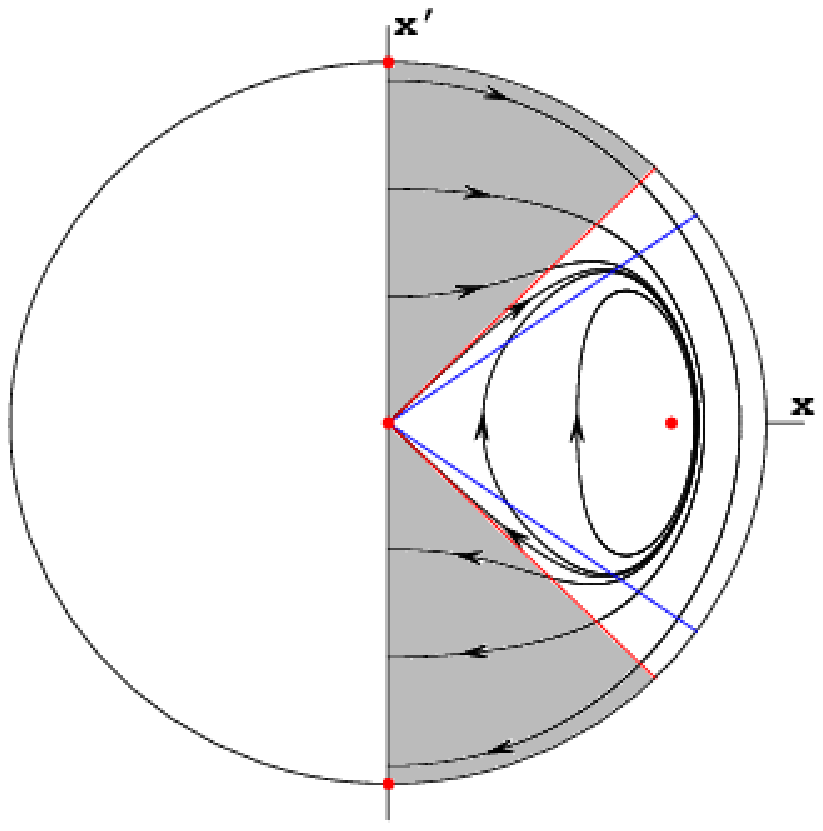}
b)\includegraphics[scale=0.9]{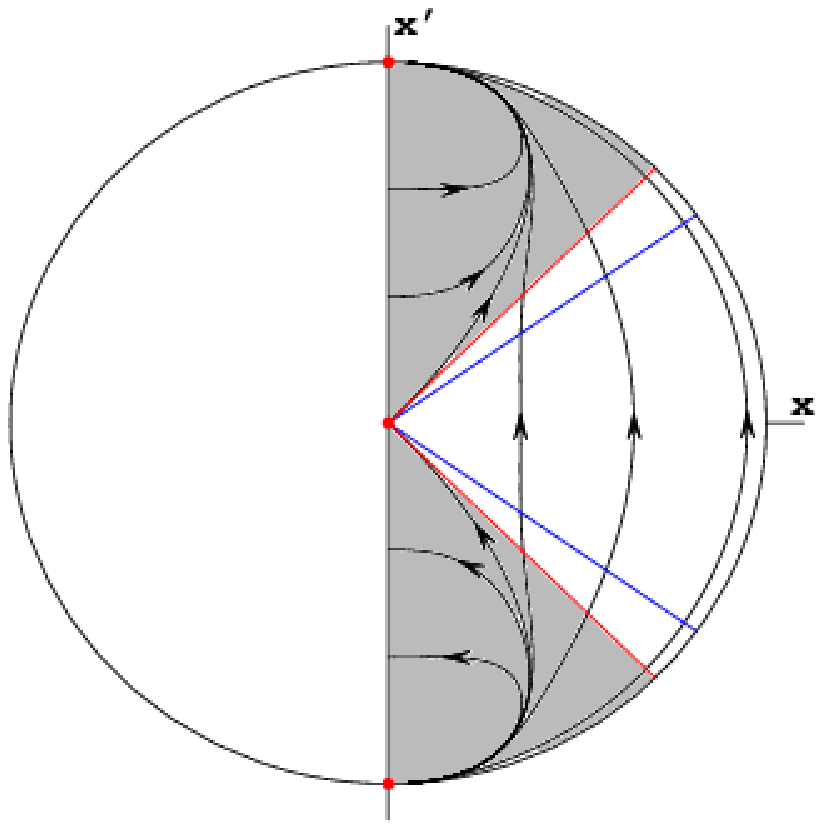}
c)\includegraphics[scale=0.9]{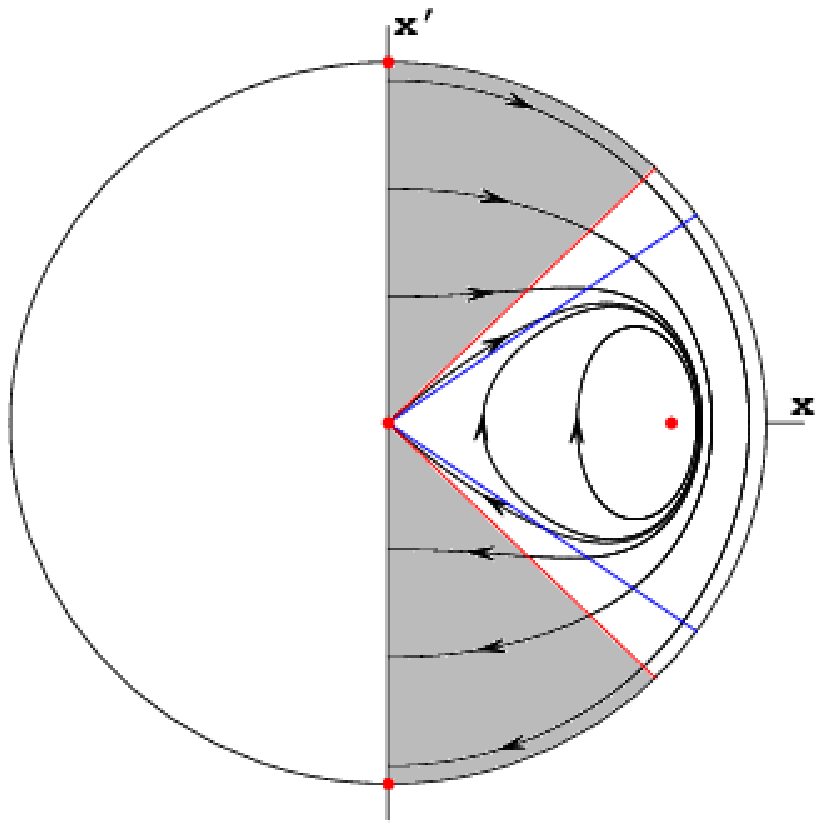}
d)\includegraphics[scale=0.9]{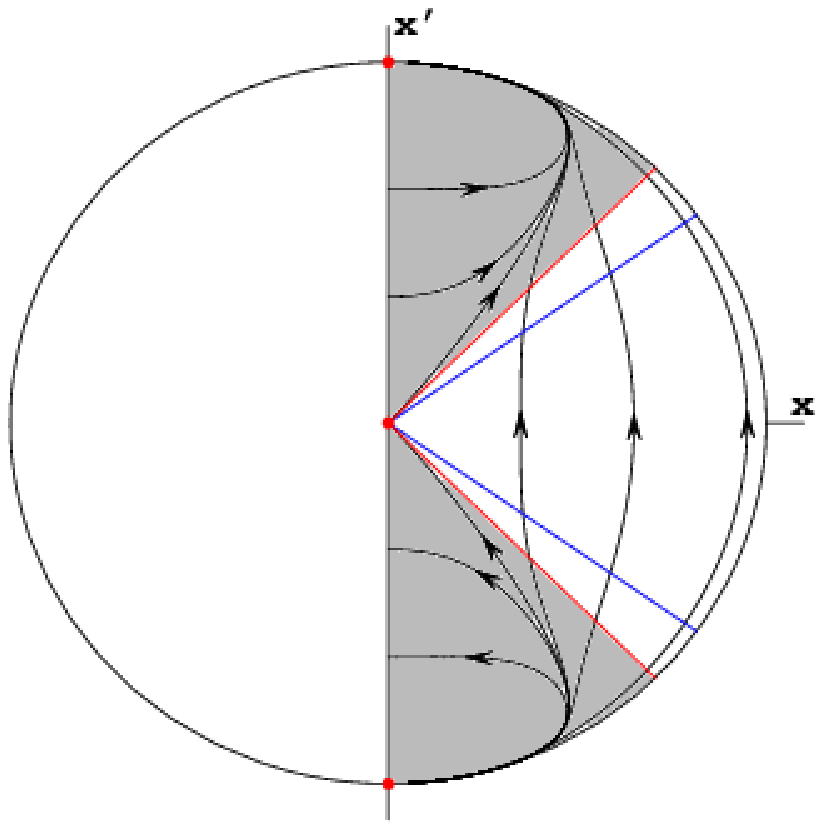}
\caption{Phase portraits for a) $\Lambda>0$ and  $n=0$; b) $\Lambda<0$ and 
$n=0$; c) $\Lambda>0$ and $n=-1/4$; d) $\Lambda<0$ and $n=-1/4$ in the case A
type modifications. Note that 
the cases $n=0$ and $n=-1/4$ are topological equivalent.}
\label{fig:4}
\end{figure}

In Figs~\ref{fig:1},~\ref{fig:2},~\ref{fig:3} and~\ref{fig:4}
we present the phase portraits for various values of the cosmological constant
$\Lambda$ and the parameter $n$ for the model with modification $\mathcal{O}(\mu^4)$.

In (\ref{timea}) the term under square root needs to be positive. In our new
variables $(x,x')$ the condition $\rho \ge 0$ (\ref{rho}) corresponds to
$$
\frac{1}{\Lambda_{0}}\bigg(\alpha+\frac{\Lambda}{3}
x^{\frac{3}{1-n}}\bigg)=x^{2}-x'^{2} \ge 0.
$$
The physical region in the phase space is defined by the condition
$x'^{2} \le x^{2}$. In the phase portraits this region is labelled as white. 
The time transformation (\ref{timea}) is singular at 
$1-\frac{2}{3}\frac{\rho}{\rho_{c}}=0$ which, in the new variables, 
corresponds to $x'^{2}=\frac{1}{2}x^{2}$. Location of such points in the phase
plane denote curvature singularities of the original system ($H^{2}=\infty$ see
(\ref{Fried1})) and are labelled as a solid blue line (electronic version).

\subsection{Quadratic modifications from $\mathcal{O}(\mu^{4})$}

The employ of the quadratic contribution from the $\mathcal{O}(\mu^{4})$ studied in the
section \ref{sec:2}  lead to the phenomenological Hamiltonian in the form 
\begin{equation}
\mathcal{H}_{\text{phen}} = -\frac{3}{\kappa\gamma^{2}\bar{\mu}^{2}} 
\Big(\sin^{2}{(\bar{\mu}c)} +\frac{1}{3}\sin^{4}{(\bar{\mu}c)}\Big) + |p|^{3/2}\rho
\end{equation}
where matter content is the same as in the previous case and is given by (\ref{rho}).
In the same way like in the previous case we aim to calculate the effective Friedmann 
equation and then rewrite it in the point-like particle form. To achieve it we derive 
\begin{equation}
\dot{p} = \left\{p,\mathcal{H}_{\text{phen}}  \right\} = \frac{2\sqrt{|p|}}{\gamma\bar{\mu}}
\sin{(\bar{\mu}c)}\cos{(\bar{\mu}c)} 
\big(1 + \frac{2}{3}\sin^{2}{(\bar{\mu}c)}\big)
\end{equation}
and with use of the Hamiltonian constraint we obtain the modified Friedmann equation  
\begin{equation}
H^2 = \frac{\kappa}{3} \rho \left( 1+\frac{4}{3} \frac{\rho}{\rho_{\text{c}}} \right) 
 \frac{   5-3\sqrt{1+\frac{4}{3} \frac{\rho}{\rho_{\text{c}}} } }{1+\sqrt{1+\frac{4}{3} \frac{\rho}{\rho_{\text{c}}} }}.
\label{Fried2}
\end{equation}
Changing of variable $x=p^{1-n}$ we rewrite equation (\ref{Fried2}) to the form
\begin{equation}
\frac{\dot{x}^2}{2} \frac{1}{(1-n)^2}\frac{3}{\kappa \rho}
\frac{1+\sqrt{1+\frac{4}{3} \frac{\rho}{\rho_{\text{c}}} }}{ 5-3\sqrt{1+\frac{4}{3} \frac{\rho}{\rho_{\text{c}}} }}
-\frac{1}{2}x^{2}\Big(1+\frac{4}{3}\frac{\Lambda}{\Lambda_0}
x^{\frac{1+2n}{1-n}}\Big)
 =2\frac{\alpha}{\Lambda_0}. 
\end{equation}
Performing the time transformation 
\begin{equation}
\frac{d}{d\tau} =  \sqrt{\frac{1}{(1-n)^2}\frac{3}{\kappa \rho}
\frac{1+\sqrt{1+\frac{4}{3} \frac{\rho}{\rho_{\text{c}}} }}{ 5-3\sqrt{1+\frac{4}{3} \frac{\rho}{\rho_{\text{c}}} }} }  
 \frac{d}{dt}
\label{timeb}
\end{equation}
we obtain Hamiltonian of the point-like particle of the unit mass
\begin{equation}
\frac{1}{2}x'^{2} + V(x) = E
\label{newtb}
\end{equation}
where the potential function is given by
\begin{equation}
V(x) = -\frac{1}{2}x^{2}\Big(1+\frac{4}{3}\frac{\Lambda}{\Lambda_0}
x^{\frac{1+2n}{1-n}}\Big)
\label{potb}
\end{equation}
and total energy
\begin{equation}
E = 2\frac{\alpha}{\Lambda_0}.
\end{equation} 

\begin{figure}
\includegraphics[scale=1]{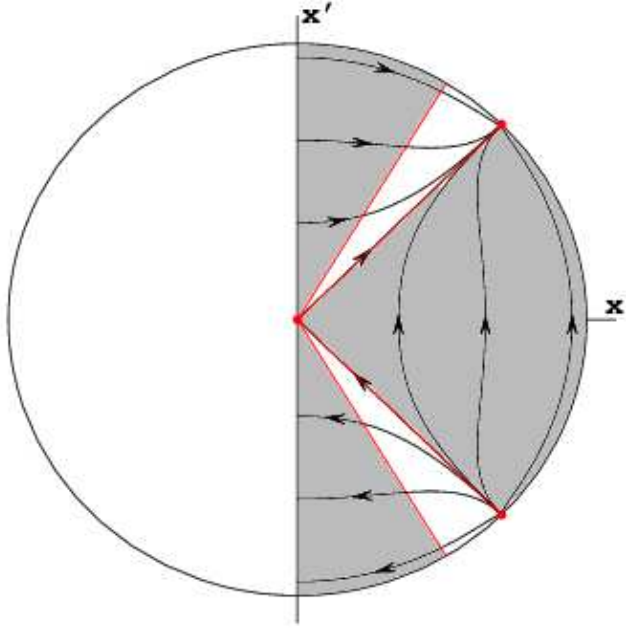}
\caption{The phase portrait for $\Lambda=0$, arbitrary $n$ and the type B 
quadratic modifications. The shaded regions are physically not allowed. The 
borders of these regions denote the phase space location of the state $H^{2}=0$, 
only the scalar field ($\alpha>0$) trajectories are located in the physical 
domain. For $x'>0$ they start from the state $H=0$ and tend to the final state 
of $H=0$ at $x=\infty$. The states $H=0$ represent the static Einstein 
universes which are reached asymptotically.}
\label{fig:5}
\end{figure}

\begin{figure}
a)\includegraphics[scale=1]{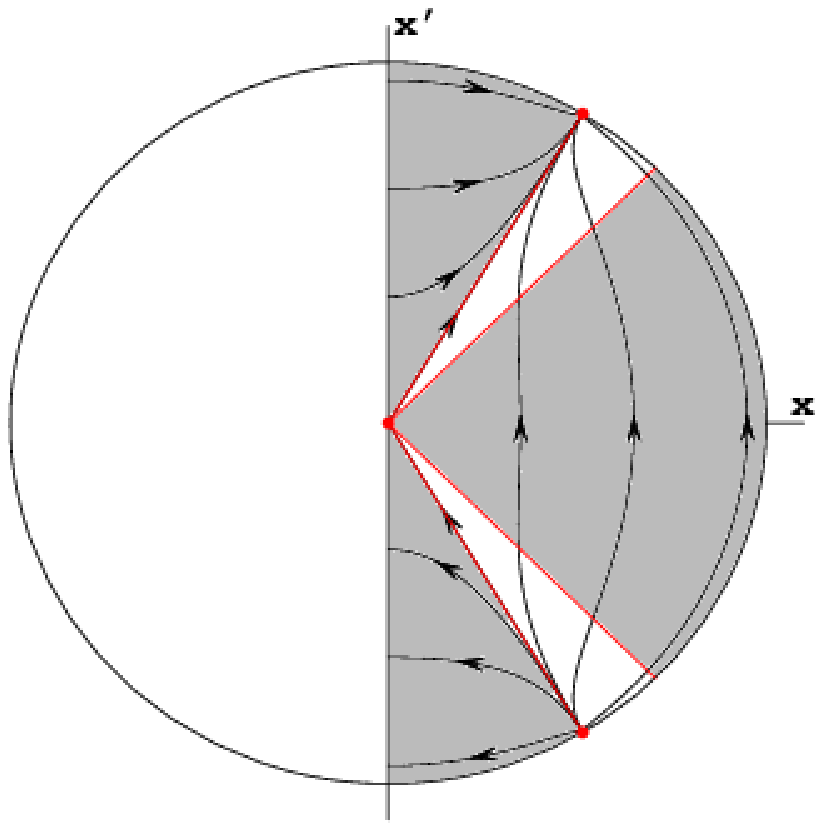}
b)\includegraphics[scale=1]{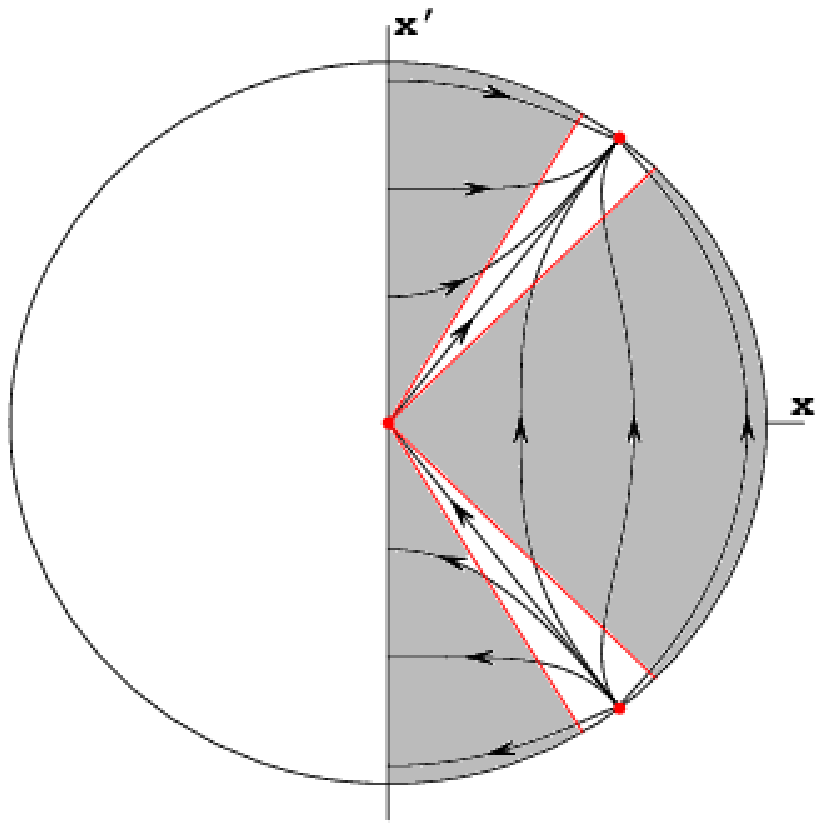}
\caption{The phase portraits for $n=-1/2$ and the type B quadratic modifications:
a) For $\frac{\Lambda}{\Lambda_0}=\frac{4}{3}$ only $\alpha<0$ trajectories are physical
and for $x'>0$ they start form the $H^{2}=0$ state for finite $x$ and go to
$H^{2}=0$ for $x=\infty$; b) For $0<\frac{\Lambda}{\Lambda_0}<\frac{4}{3}$ both types of
trajectories are allowed and they start form $H^{2}=0$ with finite $x$ and tend
to the de Sitter state at infinity.}
\label{fig:6}
\end{figure}

\begin{figure}
a)\includegraphics[scale=1]{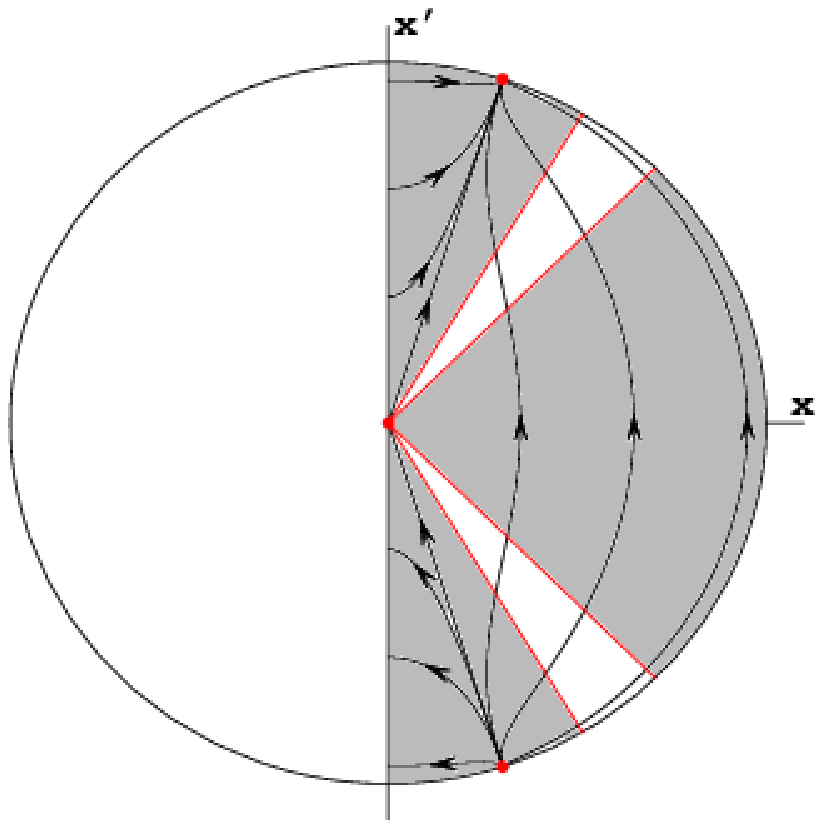}
b)\includegraphics[scale=1]{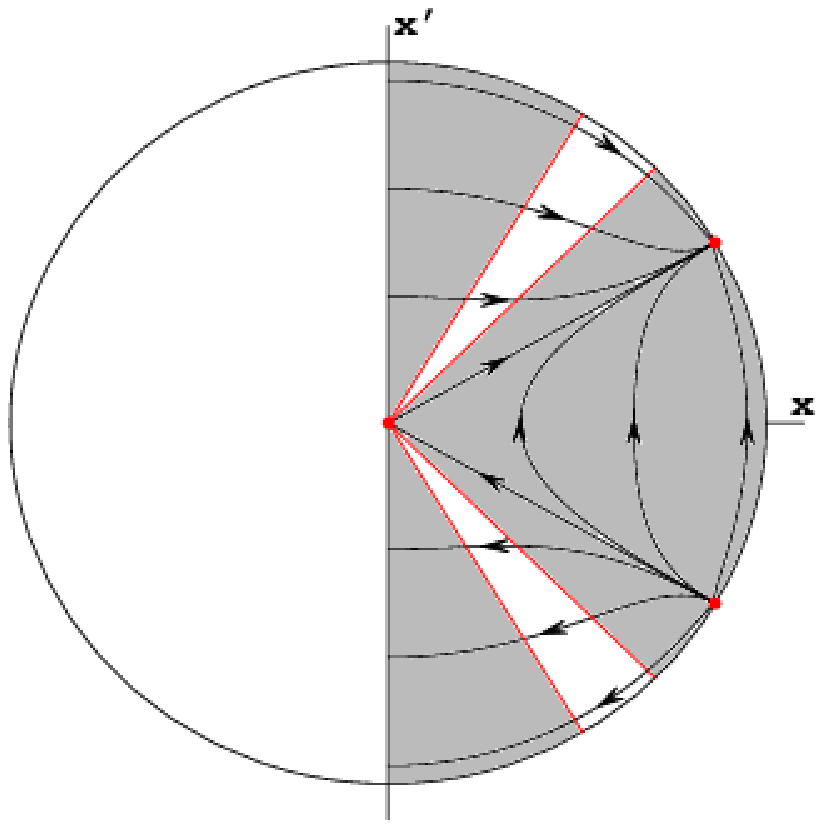}
\caption{The phase portraits for $n=-1/2$ and the type B quadratic modifications
from $\mathcal{O}(\mu^{4})$:
a) $\frac{\Lambda}{\Lambda_0}>\frac{4}{3}$, b) $-\frac{3}{4}<\frac{\Lambda}{\Lambda_0}<0$.}
\label{fig:7}
\end{figure}

\begin{figure}
a)\includegraphics[scale=0.9]{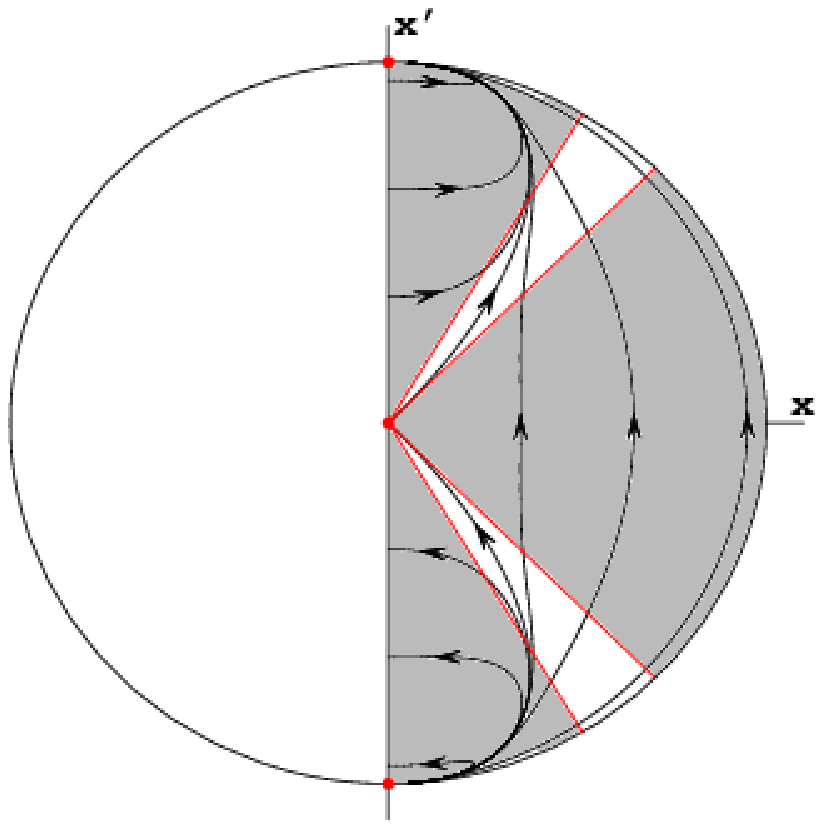}
b)\includegraphics[scale=0.9]{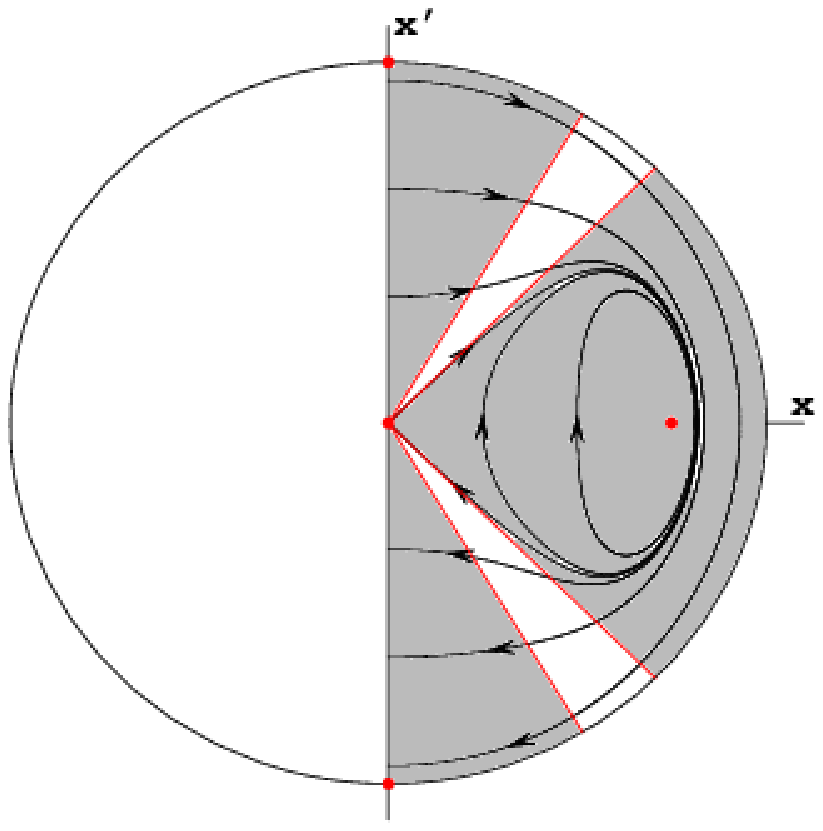}
c)\includegraphics[scale=0.9]{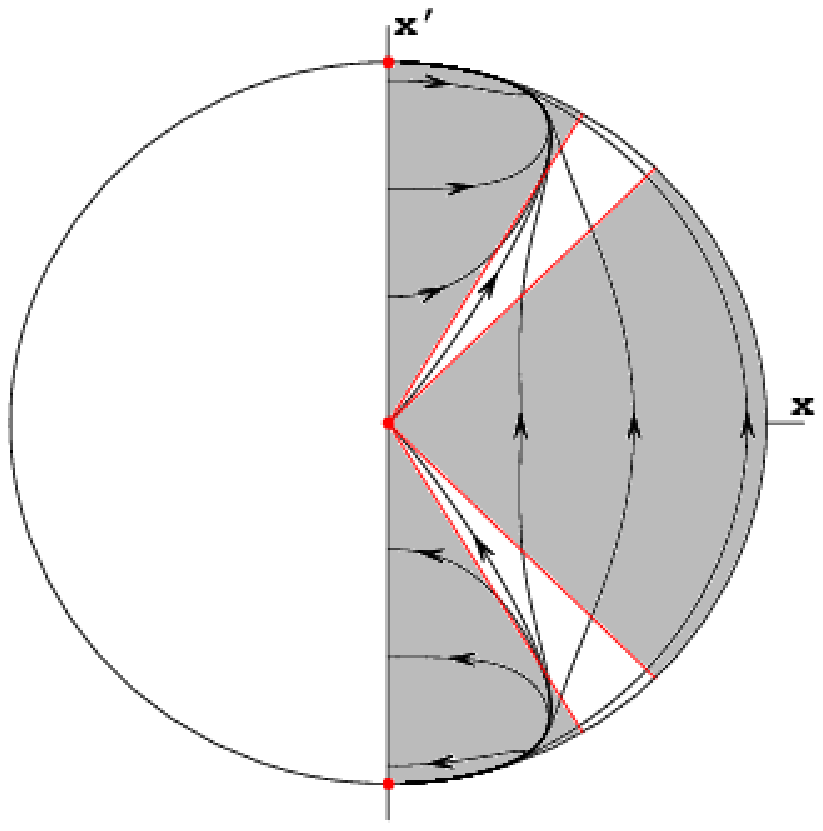}
d)\includegraphics[scale=0.9]{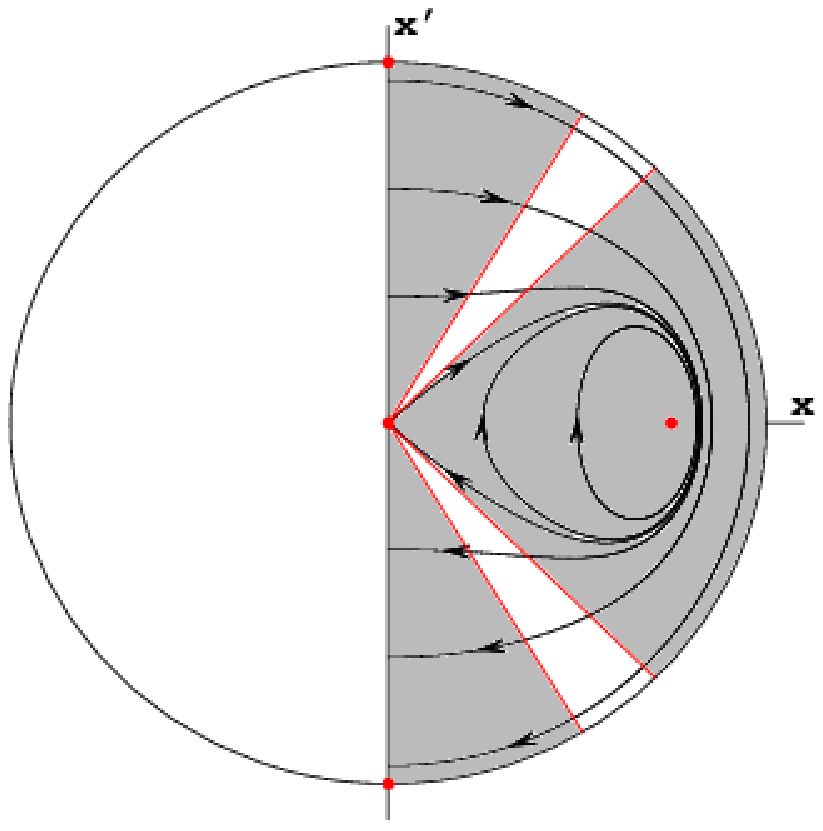}
\caption{The phase portraits for model with the type B quadratic
modifications for  $n=0$: a) and b), $n=-1/4$: c) and d) and $\Lambda>0$: a) and
c) (left panel), $\Lambda<0$: b) and d) (right panel).
The physical trajectories lie in the domain labelled as white.}
\label{fig:8}
\end{figure}

Figs~\ref{fig:5},~\ref{fig:6},~\ref{fig:7} and~\ref{fig:8} represent the phase 
portraits for model with this type modification. 

For this type of corrections the condition $\rho \ge 0$ corresponds to $x'^{2}\ge x^{2}$.
There is another condition for positivity of the Hubble function (\ref{Fried2})
and the time transformation (\ref{timeb}) which is
$\frac{5}{3} \ge \sqrt{1+\frac{4}{3}\frac{\rho}{\rho}_{c}}$. In the new variables
$(x,x')$ this is $x'^{2} \le \frac{25}{9}x^{2}$. The physical domain in the phase
space for the type B corrections is defined by 
$$
x^{2}\le x'^{2} \le \frac{25}{9}x^{2}.
$$
This region is labelled as white on the phase portraits.

\subsection{Exact solutions for the type A and B modifications}

It is interesting that for both types of modifications we can find exact
solutions for $x(\tau)$ or $p(\tau)$ for $n=-1/2$ and $n=0$. In the case of
$n=-1/2$, $\alpha>0$ the modified Friedmann equation for the type A modification 
takes the form
\begin{equation}
\left( \frac{dx}{d\tau} \right)^2 = -\frac{\alpha}{\Lambda_0}+x^2
\left(1-\frac{1}{3} \frac{\Lambda}{\Lambda_0}\right). 
\label{ESeq1}
\end{equation}
The physical solutions of this equation correspond to the case
\begin{equation}
\frac{\Lambda}{\Lambda_0} < 3.
\end{equation}
Performing the direct integration of the equation (\ref{ESeq1}) we obtain solution
\begin{equation}
x(\tau) =  \sqrt{ \frac{3\alpha}{3\Lambda_0-\Lambda}} 
\cosh \left[\sqrt{1-\frac{1}{3} \frac{\Lambda}{\Lambda_0}} (\tau-\tau_0 )\right]
\end{equation}
or
\begin{equation}
p(\tau) =  \left( \frac{3\alpha}{3\Lambda_0-\Lambda} \right)^{1/3}
\cosh^{2/3} \left[\sqrt{1-\frac{1}{3} \frac{\Lambda}{\Lambda_0}} (\tau-\tau_0)\right]
\end{equation}
for the canonical variable.

In the case of $n=-1/2$, $\alpha<0$ (phantom) physical solutions are permitted
for $\Lambda\in \mathbb{R}$. For $\Lambda/\Lambda_0< 3$ we obtain
\begin{equation}
x(\tau) =  \sqrt{ \frac{-3\alpha}{3\Lambda_0-\Lambda}} 
\cosh \left[\sqrt{1-\frac{1}{3} \frac{\Lambda}{\Lambda_0}} (\tau-\tau_0)\right],
\end{equation}
for $\Lambda/\Lambda_0=3$ we obtain
\begin{equation}
x(\tau) = x_0 + \sqrt{-\frac{\alpha}{\Lambda_0}} \tau,
\end{equation}
for $\Lambda/\Lambda_0> 3$ we obtain
\begin{equation}
x(\tau) = \sqrt{ \frac{-3\alpha}{\Lambda-3\Lambda_0}}
\sin \left[\sqrt{\frac{1}{3} \frac{\Lambda}{\Lambda_0}-1} (\tau-\tau_0)\right].
\end{equation}
Solutions for $p(\tau)$ are obtained taking $p(\tau)=x^{2/3}(\tau)$.

For $n=0$ the modified Friedmann equation takes the form
\begin{equation}
\left( \frac{dx}{d\tau} \right)^2 = -\frac{\alpha}{\Lambda_0}+x^2-\frac{1}{3}
\frac{\Lambda}{\Lambda_0}x^3.
\label{ESeq2}
\end{equation}
The solution of this equation is expressed in terms of the Weierstrass
$\wp$-function
\begin{equation}
x(\tau) = \frac{\Lambda_0}{\Lambda} \left[1-12 \wp (\tau-\tau_0;g_2,g_3)
\right]
\end{equation}
where
\begin{eqnarray}
g_2 &=& \frac{1}{12}, \\
g_3 &=& -\frac{1}{216}+\frac{1}{144}\left( \frac{\Lambda}{\Lambda_0}  \right)^2
\frac{\alpha}{\Lambda_0}.
\end{eqnarray}
For the $\Lambda<0, \alpha>0$ this solution represents a bouncing type of
evolution. In the case $\Lambda>0, \alpha>0$ physical solutions are present 
when additional relation
\begin{equation}
\frac{4}{3}  >  \frac{\alpha}{\Lambda_0}\left( \frac{\Lambda}{\Lambda_0}
\right)^2
\end{equation}
is fulfilled and represents an oscillating type of evolution.
In this case solutions for $x(\tau)$ and $p(\tau)$ are equal, $p(\tau)=x(\tau)$.

Analogous solutions for the type B modifications can be also obtained with the help
of transformation $\Lambda\to-4\Lambda$ and $\alpha\to-4\alpha$ (see Table
\ref{tab:1}).

\section{General properties of the phase portraits}

The dynamical behaviour of the models considered is fully determined by the
single potential function (\ref{pota}) or (\ref{potb}). Its diagram
characterizes a type and location of the critical points of the dynamical system
\begin{equation} \label{eq:newt}
\left\{ \begin{array}{l}
x'=y \\
y'=-\frac{\partial V}{\partial x}
\end{array} \right.
\end{equation}
System~(\ref{eq:newt}) is of the Newtonian type and at the finite domain of the 
phase space there can be only a saddle or a centre type critical points (or 
degenerated which are not interesting because we are only looking for generic 
cases). The type of a critical point is determined from the characteristic 
equation of the linearization matrix, namely
\begin{equation}
\lambda^{2} = - \frac{\partial^{2}V}{\partial x^{2}}\bigg|_{x_{0}},
\end{equation}
calculated at the critical point. If $-\frac{\partial^{2}V}{\partial
x^{2}}\big|_{x_{0}}>0$ eigenvalues are real of opposite signs and we have a 
saddle type critical point, and in case if
$-\frac{\partial^{2}V}{\partial x^{2}}\big|_{x_{0}}<0$ eigenvalues are purely
imaginary of opposite signs and a critical point is of a centre type.
\begin{table}
\caption{The Hamiltonian and derivative of the potential at the critical point for 
both types of corrections} \label{tab:1}
\begin{tabular}{|c|c|c|}
\hline
 &  A & B \\
   \hline
 Hamiltonian & $\mathcal{H}_{A}=\frac{1}{2}x'^{2}+V(x)=-\frac{1}{2}\frac{\alpha}{\Lambda_0}$
 & $\mathcal{H}_{B}=\frac{1}{2}x'^{2}+V(x)=2 \frac{\alpha}{\Lambda_0}$ \\
 potential &  $V(x)=-\frac{1}{2}x^{2}\bigg(1-\frac{1}{3}\frac{\Lambda}{\Lambda_0} 
   x^{\frac{1+2n}{1-n}}\bigg)$ &
   $V(x)=-\frac{1}{2}x^{2}\bigg(1+\frac{4}{3}\frac{\Lambda}{\Lambda_0}
   x^{\frac{1+2n}{1-n}}\bigg)$ \\
\hline
 critical points :  & & \\
 arbitrary $n$ & $x_{0}=0$, $\quad -\frac{\partial^{2}V}{\partial
 x^{2}}\Big|_{x_{0}}=1$ &  $x_{0}=0$, $\quad -\frac{\partial^{2}V}{\partial
  x^{2}}\Big|_{x_{0}}=1$\\
 $-1/2<n\le0$ & $x_{0}^{\frac{1+2n}{1-n}}=2(1-n)\frac{\Lambda_{0}}{\Lambda}$,
 $\Lambda>0$,   &
$x_{0}^{\frac{1+2n}{1-n}}=-\frac{1}{2}(1-n)\frac{\Lambda_{0}}{\Lambda}$,
$\Lambda<0$, \\ 
 &  $-\frac{\partial^{2}V}{\partial x^{2}}\Big|_{x_{0}}=-\frac{1+2n}{1-n}$ &
 $-\frac{\partial^{2}V}{\partial x^{2}}\Big|_{x_{0}}=-\frac{1+2n}{1-n}$ \\
 \hline
\end{tabular}
\end{table}

For the model with type A modifications (see also Table \ref{tab:1})
\begin{itemize}
\item{$n=-1/2$, $x_{0}=0$:
$\lambda=\pm\sqrt{1-\frac{1}{3}\frac{\Lambda}{\Lambda_0}} $}
\subitem{a saddle for $ \frac{\Lambda}{\Lambda_0} < 3 $}
\subitem{a centre for $ \frac{\Lambda}{\Lambda_0} > 3 $}
\item{$-1/2<n\le0$:}
\subitem{$x_{0}=0$ : $\lambda=\pm1$ a saddle} 
\subitem{$x_{0}^{\frac{1+2n}{1-n}}=2(1-n)\frac{\Lambda_0}{\Lambda}$ and
$\Lambda>0$: $\lambda=\pm i \sqrt{\frac{1+2n}{1-n}}$ a centre}
\end{itemize}

For the model with type B modifications (see also Table \ref{tab:1})
\begin{itemize}
\item{$n=-1/2$, $x_{0}=0$:
$\lambda=\pm\sqrt{1+\frac{4}{3}\frac{\Lambda}{\Lambda_0} }$}
\subitem{a saddle for $\frac{\Lambda}{\Lambda_0} >-\frac{3}{4}$}
\subitem{a centre for $\frac{\Lambda}{\Lambda_0} <-\frac{3}{4}$}
\item{$-1/2<n\le0$:}
\subitem{$x_{0}=0$ : $\lambda=\pm1$ a saddle}
\subitem{$x_{0}^{\frac{1+2n}{1-n}}=-\frac{1}{2}(1-n)\frac{\Lambda_0}{\Lambda}$
and $\Lambda<0$: $\lambda=\pm i \sqrt{\frac{1+2n}{1-n}}$ a centre}
\end{itemize}

Therefore all information about the critical points and their type contains 
the single potential function. 

Figs~\ref{fig:1}--\ref{fig:4} show representative cases of the global dynamics 
of the system of equations (\ref{newta}) and (\ref{pota}). 
Figs~\ref{fig:5}--\ref{fig:8} show representative cases of the global dynamics 
for type B corrections. 

We obtain different phase portraits for basic model parameters $(n,\Lambda)$, where
$-1/2 \le n \le 0$ and the cosmological constant belongs to the some
characteristic intervals. For comparison we also present phase portraits for the
case of the vanishing cosmological constant Fig.~\ref{fig:1} for the case A and
Fig.~\ref{fig:5} for the B case. In the case A we can 
observe only critical point of a centre type located at the $x$ axis and
critical point of a saddle type at the origin of the phase space (in principle
at the origin a centre type critical point for $\Lambda/\Lambda_{0}>3$ can be
present, but this case in not interesting because of it's structural instability). Along the
trajectories it is measured the new time variable which is a monotonic function
of the original cosmological time but this flow can be incomplete, i.e.,
infinite cosmological time is mapped on the finite interval measured along the
trajectory. 
 
\begin{figure}[ht!]
\centering
\includegraphics[width=9cm,angle=0]{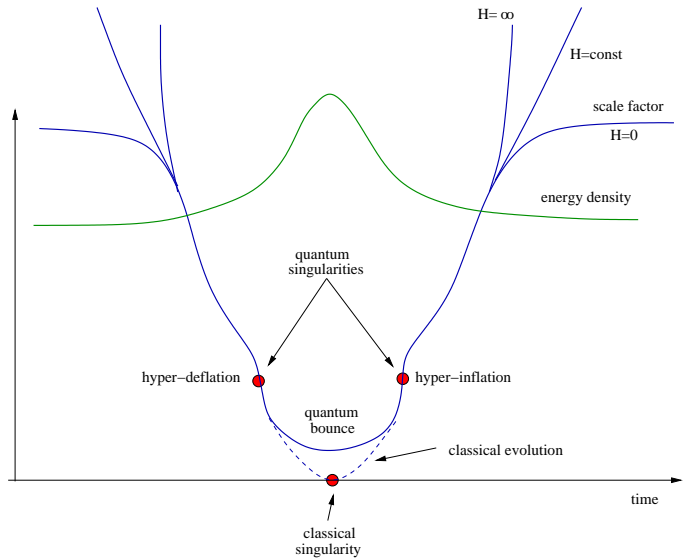}
\caption{The schematic picture of the evolution with the positive $\Lambda$ for 
the monotonic models. The green curve represents the energy density. It is 
worth to note that this energy density is finite during whole evolution, even 
during transitions through singularities. The blue curve represents the 
schematic evolution of the scale factor for the investigated model. The dashed 
curve represents the classical evolution which is not realized in the presented 
quantum model.}
\label{bounce}
\end{figure}

\begin{figure}[ht!]
\centering
\includegraphics[width=12cm,angle=0]{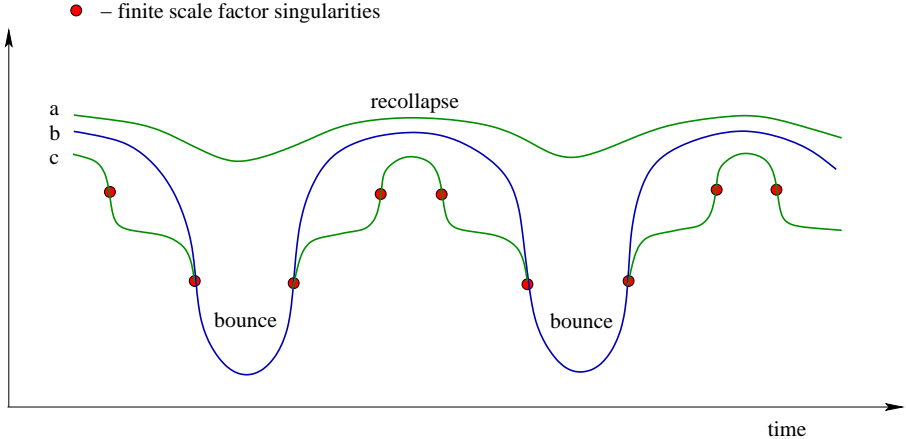}
\caption{Generic evolutional scenarios for bouncing-oscillatory models with $\Lambda>0$
and the value of critical energy constant is
$E_{\text{c}}=-\frac{1}{12}(1+2n)\Big((1-n)\frac{\Lambda}{\Lambda_{0}}\Big)^{2\frac{1-n}{1+2n}}$:
a) the oscillating model without the singularities $(E<E_{\text{c}})$; 
b) the oscillating model with double singularities at finite scale factor per period $(E=E_{\text{c}})$; 
c) the oscillating model with four singularities at the finite scale factor per period $(E>E_{\text{c}})$.}
\label{oscil}
\end{figure}

In the case of the positive cosmological constant and the case A modifications
two characteristic evolutional scenarios are realized: 
\begin{itemize}
\item{A trajectory starts from the contracting deS$_{-}$ stage and finishes at
the expanding deS$_+$ stage. In the middle state the trajectory is undergoing
the bounce in the vicinity of the intersection with the $p$-axis. In other words
for such a evolution deS$_-$ is the initial state and deS$_+$ is the final state
(see Fig.~\ref{bounce}).}
\item{There is new non expected behaviour for the positive cosmological
constant -- emerging of oscillatory behaviour -- without initial and final
singularities and, in special case, without hyper-inflation and hyper-deflation
states (see Fig.~\ref{oscil}) for
$$\alpha=\frac{\kappa}{6}p_{\phi}^{2} >
\frac{1}{6}(1+2n)\Big((1-n)\frac{\Lambda}{\Lambda_{0}}\Big)^{2\frac{1-n}{1+2n}}.$$}
\end{itemize}

What it is interesting in these scenario is the universe is undergoing a phase
of hyper-inflation in the vicinity of a singularity. This singularity is a
curvature singularity at which the scale factor (as well as density) is finite
but the Hubble function assumes infinite value. We mark a blue line in the phase
portraits which shows the singularities reached by different trajectories. The
algebraic equation for this line is independent of the value of the cosmological
constant. But note that there are trajectories which do not intersect this line
at all, they intersect it only at one point and at least they intersect it at
two points. These three cases have been illustrated schematically in
Fig.~\ref{oscil}. If $E<E_{\text{c}}$ singularities are avoided. 

The new evolutional scenario offered by the quantum correction is oscillating 
behaviour in universes with double singularities during the expansion (and
contracting) period for the positive $\Lambda$. For other propositions of 
oscillatory models with bounce and phantom see \cite{Freese:2008pu}.

All emerging evolutional scenarios for the origin of the universe from the 
bounce are in principle testable because we have the Hubble function in respect 
of the redshift, which in order can be starting point for cosmography. In 
principle if the size of the universe is going to infinity there are three 
types of exits: $H$ is going to infinity, $H$ is approaching to the de Sitter 
state or $H$ is going to the zero. All these long term behaviour one can find 
in Fig.~\ref{bounce}. 

\begin{figure}
\includegraphics[scale=0.75]{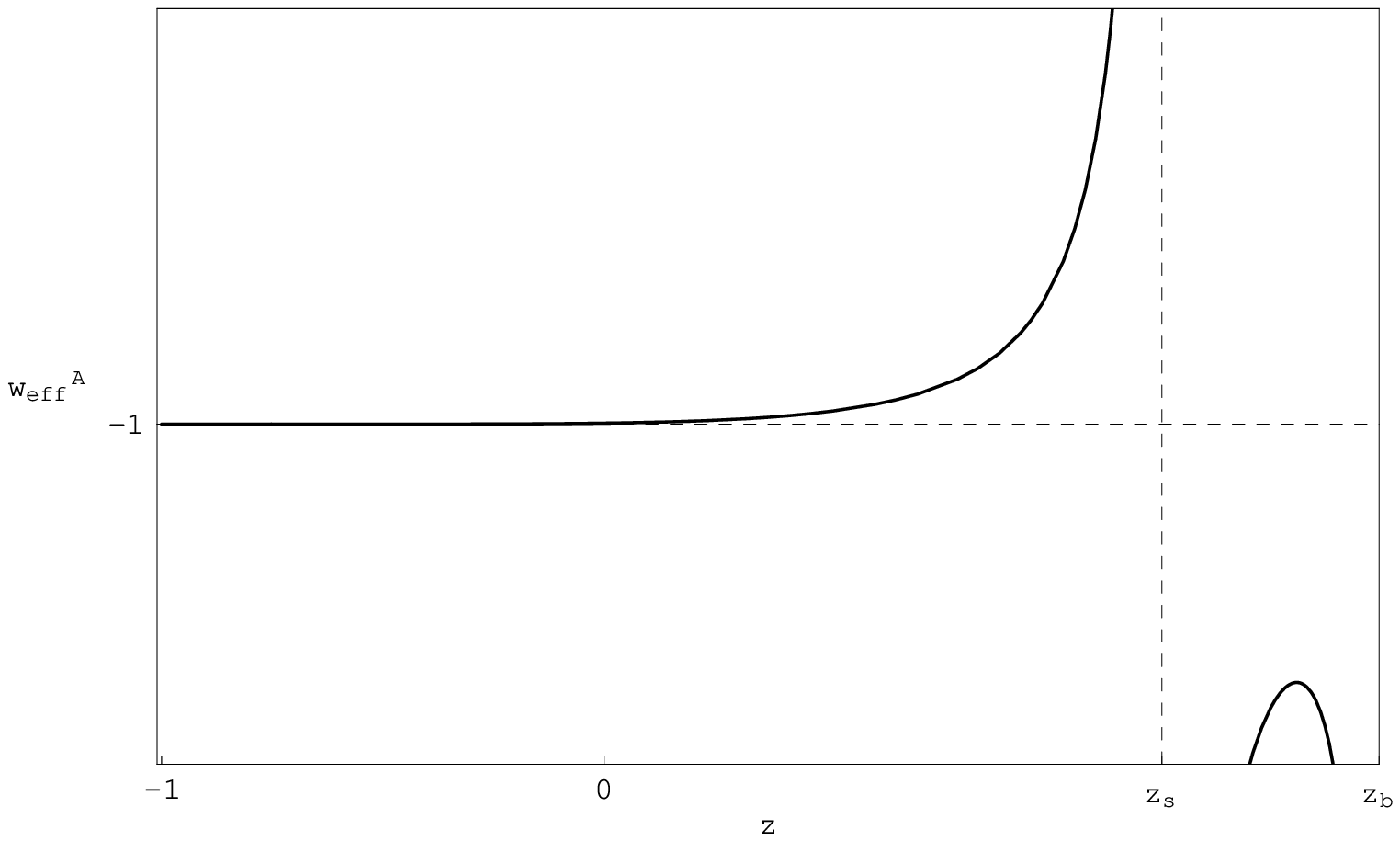}
\includegraphics[scale=0.75]{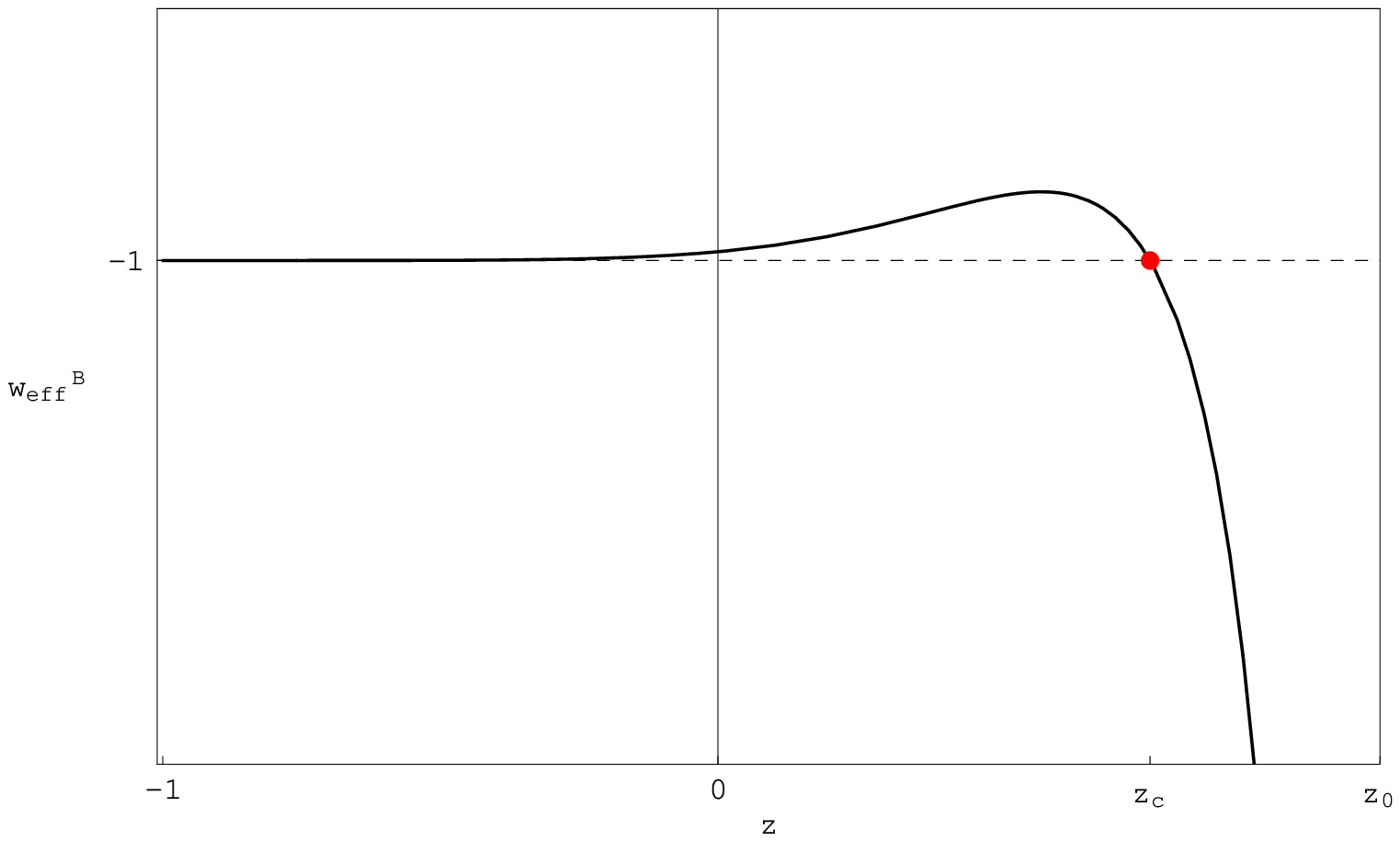}
\caption{The dependence of the effective equation of state parameter on the redshift
$z$ for both cases A (top) and B (bottom) of quantum corrections for $n=-1/2$ 
and $\Lambda>0$ (see phase portraits Fig.~\ref{fig:2}a and Fig.~\ref{fig:6}b). 
In both cases energy density approaches the cosmological constant 
$w_{\text{eff}}=-1$ case at the long time evolution. Top panel: in the type A
correction the evolution begins at the bounce for the redshift value 
$z_{\text{b}}$, next the curvature singularity is present at $z_{\text{s}}$ 
and then $w_{\text{eff}}$ approaches the $-1$ value. Bottom panel: in the type 
B correction the evolution begins at $z_{0}$ when $H^{2}=0$, next the crossing 
of the phantom divide line is present at $z_{\text{c}}$ (marked as a dot) and 
then again $w_{\text{eff}}$ approaches $-1$.}
\label{fig:weff}
\end{figure}

The quantum corrections considered are dynamically equivalent to the effects of
the decaying cosmological constant term with the effective equation of state
coefficient parameterized by the scale factor or the redshift. This fictitious
fluid which mimics dynamical effects of the quantum corrections can be
calculated in the exact form for the both types of corrections. Using the 
general formulae
$$
w_{\text{eff}}=-1-\frac{1}{3}\frac{d \ln{H^{2}}}{d \ln{a}}
$$
and modified Friedmann equations (\ref{Fried1}) and (\ref{Fried2}) we obtain
following forms of the equation of state parameter: for the type A
correction
\begin{equation}
w^{A}_{eff}=-1+2\frac{\alpha a^{-6}}{\alpha a^{-6}+\frac{\Lambda}{3}}
\bigg[\frac{1-\frac{2}{3}\frac{\rho}{\rho_{c}}}{1-\frac{\rho}{3\rho_{c}}} +
\frac{\frac{\rho}{3\rho_{c}}}{\Big(1-\frac{2}{3}\frac{\rho}{\rho_{c}}\Big)\Big(\frac{3}{4}(1-\frac{2}{3}\frac{\rho}{\rho_{c}})^{2}+\frac{1}{4}\Big)}\bigg]
\label{wa}
\end{equation}
and for the type B correction
\begin{equation}
w^{B}_{eff}=-1+2\frac{\alpha a^{-6}}{\alpha a^{-6}+\frac{\Lambda}{3}} 
\bigg[\frac{1+\frac{8}{3}\frac{\rho}{\rho_{c}}}{1+\frac{4}{3}\frac{\rho}{\rho_{c}}}
-
\frac{\frac{16}{3}\frac{\rho}{\rho_{c}}}
{\Big(5-3\sqrt{1+\frac{4}{3}\frac{\rho}{\rho_{c}}}\Big)\Big(1+\sqrt{1+\frac{4}{3}\frac{\rho}{\rho_{c}}}\Big)\sqrt{1+\frac{4}{3}\frac{\rho}{\rho_{c}}}}\bigg]
\label{wb}
\end{equation}
where
\begin{equation}
\frac{\rho}{\rho_{c}}=\frac{3}{\Lambda_{0}} a^{2(2n+1)}\Big(\alpha a^{-6} +
\frac{\Lambda}{3}\Big).
\label{rhorhoc}
\end{equation}
Equation (\ref{rhorhoc}) gives us opportunity to calculate the redshift distance
to the characteristic moments of the universe evolution. For the type A
correction the bounce occurs when $\rho/\rho_{c}=3$ and using (\ref{rhorhoc})
we have that at the scale factor value fulfilling relation
$$
a^{2(2n+1)}\Big(\frac{\alpha}{\Lambda_{0}}a^{-6}+\frac{1}{3}\frac{\Lambda}{\Lambda_{0}}\Big)=1
$$
the bounce is present. In the special case of $n=-1/2$, the value of the 
redshift at the bounce is
\begin{equation}
z_{\text{b}}=-1+\sqrt[6]{\frac{\Lambda_{0}}{\alpha}\Big(1-\frac{1}{3}\frac{\Lambda}{\Lambda_{0}}\Big)}.
\end{equation}
Another characteristic moment of the model evolution is the curvature 
singularity appearing at the redshift (for $n=-1/2$)
\begin{equation}
z_{s}=-1+\sqrt[6]{\frac{\Lambda_{0}}{\alpha}\Big(\frac{1}{2} -
\frac{1}{3}\frac{\Lambda}{\Lambda_{0}}\Big)}.
\end{equation}

For the type B corrections evolution begins when $\rho/\rho_{c}=4/3$, that is,
for $n=-1/2$ at
\begin{equation}
z_{0}=-1+\sqrt[6]{\frac{1}{3}\frac{\Lambda_{0}}{\alpha}
\Big(\frac{4}{3}-\frac{\Lambda}{\Lambda_{0}}\Big)}
\end{equation}
There is another important moment in the evolution appearing only for the type B 
correction, namely the crossing of the phantom divide line (Fig.~\ref{fig:weff} 
bottom). In this case the during the evolution of the universe $w_{\text{eff}}$ 
starts from a large negative value, crosses the line $w_{\text{eff}}=-1$, then 
reaches the maximum and now decreases to asymptotically to minus one. 
Solving Eq.~(\ref{wb}) for $w^{B}_{\text{eff}}=-1$ we find that 
$\rho/\rho_{c}=\frac{1}{8}(1+2\sqrt{6})$ and for $n=-1/2$ at the redshift value
\begin{equation}
z_{\text{c}}=-1+\sqrt[6]{\frac{1}{3}\frac{\Lambda_{0}}{\alpha}
\Big(\frac{1}{8}(1+2\sqrt{6})-\frac{\Lambda}{\Lambda_{0}}\Big)}
\end{equation}
the crossing of the phantom divide line happens.

\section{Conclusions}
\label{sec:summary}

In this paper we have considered the influence of the modifications to the 
field strength expression for the effective dynamics in the loop quantum 
cosmology. We have studied the sensitivity of evolutional scenarios on the 
choice of model parameters $(n,\Lambda)$. For the classification of evolutional 
paths in arising the loop quantum cosmology models it is natural to adopt the 
dynamical systems methods which offers full classification of dynamical outcome 
for all admissible initial conditions. We consider two types of quantum 
modifications labelled as modification (A) $\mathcal{O}(\mu^{4})$
and the quadratic modifications from $\mathcal{O}(\mu^{4})$ (B) (for
details see Section \ref{sec:2}). 

From mathematical point of view the dynamics is represented by $2D$ dynamical
system of the Newtonian type. From the cosmological point of view quantum
corrections are manifested by decaying part of the cosmological constant
$\Lambda$ term. Main advantage of using dynamical systems methods lies in the
possibility of studding complexity of dynamics. In the geometric language of the 
phase space we can analyze the asymptotic stability of the solutions. From this 
analysis we obtain that a generic feature of dynamics is emergence of the 
bounce instead of the initial singularity. We can find easily beginning and 
final states of dynamics in its long time behaviour using the phase
space compactification with the circle at infinity. For the positive cosmological
constant and (A) type corrections, in principle, there are three types of
outcomes (and incomes respectively because the dynamics is invariant under the
reflection symmetry $\dot{p}\to-\dot{p}$), namely the expanding de Sitter state
deS$_{+}$, $H=\infty$ state or the Einstein static universe $H=0$. 

The new feature of dynamics within models with (A) type corrections is
appearance of singularities at the finite scale factor and the 
super-accelerating phase of expansion in the vicinity of this singularity. 
However, we demonstrate that this singularity can be avoided for some choice 
of the critical value of total energy $E_{\text{c}}$. 

Another new phenomenon is the emergence of oscillating models for positive 
value of the cosmological constant with single or double curvature singularities
for the finite scale factor. We demonstrate that the dynamics with (A) and (B)
a type of corrections are dual and after the transformations
$\Lambda\to-4\Lambda$ 
and $\alpha\to-4\alpha$ one can obtain the evolution of models with the
massless scalar field from evolution of the phantom scalar field (formally this
means that we consider a opposite sign of total energy and the cosmological
constant). The main difference between the models with positive and negative
cosmological constant is that in the latter case incomes and outcomes always
represents the static Einstein Universe.

\begin{acknowledgments}
This work has been supported by the Marie Curie Host Fellowships for the
Transfer of Knowledge project COCOS (Contract No. MTKD-CT-2004-517186). The
authors also acknowledge cooperation in the project
PARTICLE PHYSICS AND COSMOLOGY: THE INTERFACE (Particles-Astrophysics-Cosmology
Agreement for scientific collaboration in theoretical research).
\end{acknowledgments}

\end{document}